\documentclass{article}

\usepackage[
  backend=bibtex,
  style=numeric-comp,
  bibstyle=customref,
  giveninits=true,
  maxnames=6,
  minnames=1,
  sorting=none
]{biblatex}
\usepackage{csquotes}

\usepackage[english]{babel}

\usepackage[a4paper,top=2cm,bottom=2cm,left=3cm,right=3cm,marginparwidth=1.75cm]{geometry}

\usepackage{lmodern}
\usepackage{microtype}
\usepackage{amsmath,amssymb,amsfonts,amsthm}
\usepackage{mathrsfs}
\usepackage{graphicx}
\usepackage[colorlinks=true, allcolors=blue]{hyperref}
\usepackage{tikz}
\usepackage{circuitikz}
\usetikzlibrary{arrows}
\usepackage{pgfplots}
\usepackage{booktabs}
\usepackage[ruled, vlined,linesnumbered, commentsnumbered, longend]{algorithm2e}

\usepackage{caption}
\usepackage{subcaption}
\usepackage{placeins}

\pgfplotsset{compat=1.15}

\title{A first look at Structured-Multiscale Algebraic Multigrid for Lattice Field Theory.}

\usepackage{authblk}

\author[1,2]{Pauline Schauerte}
\author[3]{Jaime~Fabián~Nieto~Castellanos}
\author[1]{Arnold Krechel}
\author[1,2]{Marc Alexander Schweitzer}
\author[3,4]{Stefan Krieg}
\affil[1]{\small\textit{Fraunhofer Institute for Algorithms and Scientific Computing SCAI,\protect\\ 53757 Sankt Augustin, Germany}}
\affil[2]{\small\textit{Institute for Numerical Simulation, Rheinische Friedrich-Wilhelms-Universität,\protect\\ 53115 Bonn, Germany}}
\affil[3]{\small\textit{Jülich Supercomputing Centre, Forschungszentrum Jülich,\protect\\ 52428 Jülich, Germany}}
\affil[4]{\small\textit{Helmholtz-Institut für Strahlen- und Kernphysik,
Rheinische Friedrich-Wilhelms-Universität Bonn,\protect\\ 53115 Bonn, Germany}}
\date{}

\begin{document}
\maketitle

\begin{abstract}
State-of-the-art solvers for the Dirac equation in Lattice QCD are based on adaptive multigrid methods. These require fine-tuning of many algorithmic parameters to achieve optimal performance. We apply a new multigrid approach to Lattice Field Theory adapted from oil-reservoir simulations: Structured-Multiscale Algebraic Multigrid (SM-AMG). This method builds compact aggregates with overlapping borders to coarsen the grid and yields accurate interpolation. A key advantage is that aggregate size is the primary tunable parameter. For our results, we used SM-AMG in an algebraic approach, called Aggregative-Multiscale AMG (AM-AMG). We benchmark the efficiency of AM-AMG against that of DD$\alpha$AMG, a successful adaptive multigrid solver which alleviates critical slowing down. The two solvers are compared within the framework of the two-flavor Schwinger model using the Wilson discretization. On fine lattices, the operation count of both methods is similar near the critical point and for large volumes, reflecting a comparable computational cost. However, the number of fine-grid iterations is larger for AM-AMG. On coarse lattices, AM-AMG encounters difficulties to remove the low modes close to the critical mass.
\end{abstract}

\section{Introduction}

The computation of quark propagators remains one of the most computationally demanding tasks in Lattice QCD. This involves the numerical solution to the Dirac equation, a linear system of the form \begin{equation}
    Dx=\psi,
    \label{eq:DiracEq}
\end{equation}where $D$ is a matrix that represents a discretization of the Dirac operator. This operator depends on the gauge degrees of freedom and the bare fermion mass. Both simulations with the Hybrid Monte Carlo (HMC) algorithm and the subsequent determination of many observables rely on solving this problem multiple times. In addition, realistic simulations require a fine lattice spacing, large volumes and tuning the quark masses to their physical value. The latter shifts the small eigenvalues of $D$ close to the origin and results in \textit{critical slowing down}. As a consequence, Krylov solvers require many iterations to invert $D$, making straightforward simulations with these methods unfeasible for physical masses and many degrees of freedom. Multilevel preconditioners alleviate this problem, although the simplest approaches still present troubles due to the random nature of $D$. 
In particular, Adaptive Multigrid \cite{Brezina2004,Brezina2006,Brannick2008,Babich2010} has emerged as the current paradigm for simulating QCD close to the physical point. By using a smoother to generate a set of test vectors, the method captures the ``near-kernel" --the subspace spanned by eigenvectors corresponding to the smallest eigenvalues. These vectors are employed to construct an interpolator that projects the problem onto a coarser grid, where the low-mode errors are more efficiently resolved. The combination of this interpolator with an adequate smoother provides a strong preconditioner able to deal with critical slowing down. Each test vector represents a ``color'' degree of freedom on the coarse grid. The number of vectors is fine-tuned for each different scenario to prevent a misuse of computing resources. For instance, for increasing volume and near physical masses, more test vectors capture the low-lying eigenspectrum better. However, an excess of these results in  a coarse grid operator which is expensive to apply.\\

Recent advancements in solver development for Lattice QCD have focused on refining the existing adaptive-multigrid approach or using neural networks to build a multilevel preconditioner. These approaches include, for example, 
refining the setup phase \cite{whyte2025}, enhancing coarsest-level solutions \cite{espinoza2023} or using gauge-equivariant layers as a basis for constructing machine learning methods \cite{Lehner2023}. 
Despite their overall success, progress on multilevel methods outside of Lattice QCD is moving forward. With this in mind, we wish to attract attention to a recent method based on Algebraic Multiscale (AMS) \cite{zhou}, namely Structured-Multiscale Algebraic Multigrid (SM-AMG) \cite{ehrmannAMAMG}. It constructs compactly shaped aggregates with different function types for the variables inside the aggregates. The overlapping borders of these, the so-called wirebaskets, allow for a more accurate interpolation, which is ideal per aggregate.\\

As a proof of concept, we use SM-AMG as a preconditioner for the Flexible Generalized Minimal Residual Method (FGMRES) to solve 
the Dirac equation within the two-flavor Schwinger model using the Wilson discretization. This model serves as standard testbed for QCD simulations due to their physical similarities but reduced 
number of degrees of freedom. We test the scalability and robustness of SM-AMG for gauge configurations generated close to the chiral limit. Furthermore, we implement Domain Decomposition Adaptive Algebraic Multigrid (DD$\alpha$AMG) \cite{Frommer2014,Frommer2013} for the two-flavor lattice Schwinger model to benchmark SM-AMG's viability to deal with critical slowing down when compared to modern solvers. \\

\subsection{Schwinger Model}
The Schwinger model \cite{Schwinger1,Schwinger2} refers to Quantum Electrodynamics in two dimensions. Some of its physical features, for instance confinement, topology and presence of anomalies, resemble those of QCD \cite{Coleman}. These properties, together with its lower dimensionality and simple gauge group structure make it a good toy model for testing algorithms. We briefly describe the Wilson-Dirac operator on the lattice for this case.\\

Let us consider a 2d discrete volume of dimensions $N_x\cdot N_t$\begin{equation}\Lambda := \lbrace \mathbf{n} = (n_t,n_x)\,\mid\, n_t = 0, 1, \dots, N_t-1; \,n_x = 0,1 \dots, N_x -1 \rbrace.
\end{equation}The Wilson formulation in the Schwinger model is defined by the operator \begin{align}
    D_{\alpha\beta}[\mathbf{n},\mathbf{m}] &= (m_0+2)\delta_{\alpha\beta}\delta_{\mathbf{n},\mathbf{m}} \notag \\
    &- \frac{1}{2} \sum_{\mu=0}^1\left[(1-\sigma_\mu)_{\alpha\beta} \,U_\mu(\mathbf{n}) \delta_{\mathbf{n}+\hat{\mu},\mathbf{m}} + (1+\sigma_\mu)_{\alpha\beta}\, U_\mu^\dagger (\mathbf{n}-\hat{\mu}) \delta_{\mathbf{n}-\hat{\mu},\mathbf{m}}  \right],
\end{align}where $\alpha,\beta\in\lbrace0,1\rbrace$ represent the spin indices; $\mathbf{n},\mathbf{m}\in\Lambda$; $U_\mu(\mathbf{n})\in\textrm{U(1)}$; $m_0$ is the bare mass parameter and $\sigma_\mu$ are the Pauli matrices. The vector $\hat{\mu}$ represents a unit vector in the direction indexed by $\mu$.  We choose the following representation for $\sigma_\mu$ \begin{equation}
    \sigma_0 = \begin{pmatrix} 0 & 1 \\ 1 & 0\end{pmatrix}, \quad \sigma_1=\begin{pmatrix} 0 & -i \\ i & 0\end{pmatrix}.
\end{equation}The analogue of the Hermitian $\gamma_5$ matrix in this case is \begin{equation}
    \sigma_3 = i\sigma_0 \sigma_1 = \begin{pmatrix} -1 & 0 \\ 0 & 1\end{pmatrix}, \quad \sigma_3^\dagger = \sigma_3, \quad \lbrace \sigma_\mu,\sigma_3\rbrace = 0.
\end{equation}The $\sigma_3$ properties imply that the Wilson-Dirac operator satisfies\begin{equation}
    D^\dagger = \sigma_3 D \sigma_3,
\end{equation}which results in eigenvalues that occur in complex conjugate pairs. We use periodic boundaries for the space direction and antiperiodic for the time direction, \textit{i.e.}\ for any field $\psi_\mu(\mathbf{n})$ on the lattice one has \begin{align}
   \psi_\mu(n_t,(N_x-1)+1)=\psi_\mu(n_t,0), \quad\psi_\mu((N_t-1) + 1,n_x) = -\psi_\mu(0,n_x). 
\end{align}\\

In Section \ref{AMG} we briefly review the basic structure of a multilevel method. In Section \ref{DDalpha} we describe the components of our DD$\alpha$AMG implementation for the Schwinger model. In Section \ref{AM-AMG} we introduce SM-AMG and its implementation in depth. Sections \ref{results} and \ref{conclusions} contain the results and conclusions.

\section{Multilevel approach}\label{AMG}
Multilevel methods solve sparse linear equation systems iteratively, by creating hierarchy levels, either with the help of underlying geometry (\textit{e.g.}\ Geometric Multigrid \cite{trottenberg2000multigrid}), or algebraically (\textit{e.g.}\ Algebraic Multigrid \cite{stubenAMG}). In general, we look at systems of the form
\begin{align}
    Au\,=\,f \quad \text{  or  }\quad \sum_{j=1}^{n} a_{ij}u_{j}\,=\,f_{i} \quad (i\,=\,1,2,\dots,n), \label{eq:problem1}
\end{align} with $A\in\mathbb{C}^{n\times n}$. \\

Most multilevel methods can be split in two phases. The \textit{setup phase}, in which the given problem (\ref{eq:problem1}) is analyzed, the coarse levels are constructed and all operators are assembled, and the \textit{solution phase}, which actually solves the problem. The first phase can be divided into the following three parts:
\begin{itemize}
    \item \textit{Coarsening}: Define a set of variables that induces the coarse-level.
    \item \textit{Construction of interpolation} $I_{H}^{h}$: The interpolation operator $I_{H}^{h}$ is constructed to transfer variables of the coarse and fine level and the restriction operator $I_{h}^{H}$ as a reverse operator.
    \item \textit{Calculation of the coarse level operator} $A_{H}$: Based on the interpolation operator and the coarse restriction operator, the coarse level operator is 
    calculated as the Galerkin product of interpolation and restriction, $A_{H}\,=\,I_{h}^{H}A_{h}I_{H}^{h}$.
\end{itemize}

The goal of the solution phase is to iteratively compute the approximate solution $u^{h}$. For that, we need to be able to calculate the defect correction vector $e^{H}$ on the coarse level, using the 
linear system  \begin{align}
    A_{H}e^{H}\,=\,I_{h}^{H}d^{h}_{i}\,=\,I_{h}^{H}(f_{h}-A_{h}u^{h}_{i}),
\end{align}
which has to be solved on the coarse level. The solution approximation is updated by the interpolated correction $I_{H}^{h}e^{H}$ with \begin{align}
    u^{h}_{i+1}\,=\, u^{h}_{i}+I_{H}^{h}e^{H}.
\end{align}
Algorithm \ref{alg:two} illustrates the solution phase for a two-level scheme, where a residual reduction or a maximum number of iterations is used as stopping or convergence criterion. \\

In Sections \ref{DDalpha} and \ref{AM-AMG}, we explain DD$\alpha$AMG and SM-AMG in more detail. Note that both methods follow Algorithm \ref{alg:two}, but have different interpolation operators, which are used to construct the coarse level. The interpolator is used in line 5 in Algorithm \ref{alg:two} to calculate the coarse level operator $A_{H}\,=\,I_{h}^{H}A_{h}I_{H}^{h}$.

\begin{algorithm}[h]
\SetKwInOut{KwIn}{Input}
\SetKwInOut{KwOut}{Data}
\KwIn{$A=A_{f}$: fine-level matrix\\
        $f$: right-hand vector\\
        $u_{0}$: start vector\\
        $\mathscr{C}(u_{i}^{1})$: stopping/convergence criterion depending on $u_{i}$}
\KwOut{ $S$: smoothing operator on level k\\
        $I_{H}^{h}$: interpolation operator\\
        $I_{h}^{H}$: restriction operator\\
        $A_{H}$: coarse level matrix}
\While{not $\mathscr{C}(u_{i}^{1})$}
{ Apply pre-smoothing: $\tilde{u_{i}}=Su_{i-1}$\\
Calculate residual: $d_{i}^{h}=f-A_{h}\tilde{u}_{i}$\\
Restrict residual: $d_{i}^{H}=I_{h}^{H}d_{i}^{h}$\\
Solve coarse-level equation: $e_{i}^{H}=A_{H}^{-1}d_{i}^{H}$\\
Interpolate coarse-level correction: $e_{i}^{h}=I_{H}^{h}e_{i}^{H}$\\
Add the correction: $\hat{u}_{i}=\tilde{u}_{i}+e_{i}^{h}$\\
Apply post-smoothing: $u_{i}=S_{k}\hat{u}_{i}$\\
Increment $i$
}
\caption{Two-level scheme of the solution phase of AMG}
\label{alg:two}
\end{algorithm}

\section{Domain Decomposition Adaptive AMG}\label{DDalpha}

Domain Decomposition Adaptive Algebraic Multigrid (DD$\alpha$AMG) provides an efficient multilevel framework for solving eq.\ \eqref{eq:DiracEq}. The core efficiency of DD$\alpha$AMG lies in its interpolation scheme. This follows an adaptivity approach, where a set of $N_v$ test vectors  $\lbrace w_i| i=1,\cdots,N_v\rbrace$ is generated during a setup phase to capture the near-kernel of the Dirac matrix. The test vectors show the property of ``local coherence" \cite{Luescher2007}. This key observation posits that elements of the near-kernel approximately coincide across different lattice blocks. Consequently, one can represent a large portion of the near-kernel with a reduced set of test vectors decomposed over different blocks. Therefore, local coherence motivates the construction of an aggregation scheme to split the test vectors across the lattice. The resulting coarse-grid operator is then defined via the Galerkin product $D_c=P^\dagger D P$. \\

Domain decomposition plays a central role. In particular, we use a block Gauß-Seidel method as smoother, known in the Lattice QCD community as the Schwarz Alternating Procedure (SAP), see  Ref.\ \cite{Luescher2004}. This technique effectively reduces high-frequency error components through local solves on these blocks.\\

Then, an initial set of test vectors is generated by applying the smoother. These vectors are intended to approximate solutions of the homogeneous equation $Dx = 0$, thereby exposing components in the near-kernel. By varying the initial guesses and performing a short number of SAP iterations, we obtain  $N_v$ distinct near-kernel candidate vectors, which are then used to construct the interpolator $P$.\\

\begin{figure}[h]
    \centering
    \begin{subfigure}[t]{0.5\textwidth}
        \centering
        \includegraphics[width=1\linewidth]{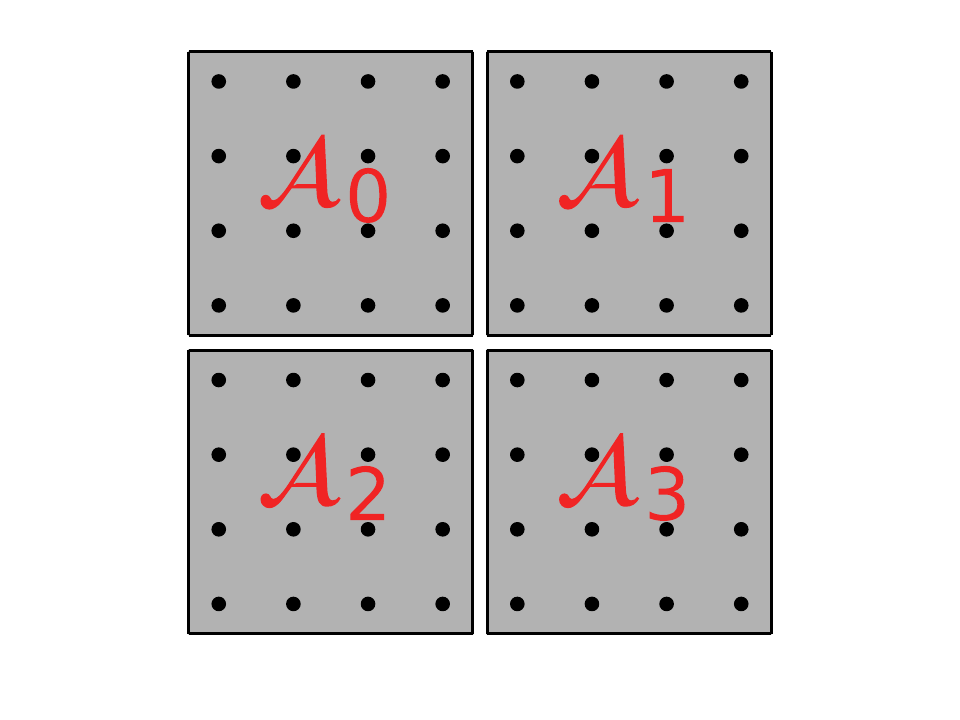} 
        \caption{Aggregation of the lattice}
    \end{subfigure}%
    ~ 
    \begin{subfigure}[t]{0.5\textwidth}
        \centering
        \includegraphics[width=1\linewidth]{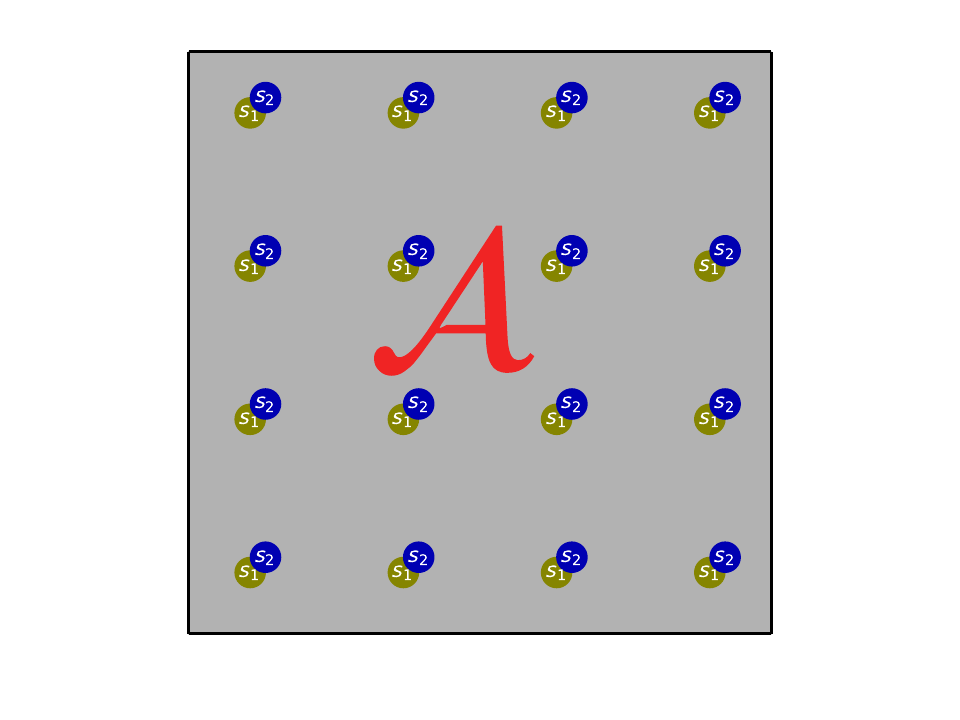}
        \caption{Inside the aggregates each site has two spin degrees of freedom, which are interpolated independently.}
    \end{subfigure}%
    \caption{We partition the lattice into equally sized groups of sites, \textit{i.e.}\ aggregates. Each lattice site carries two spin degrees of freedom, which are separated for the purpose of interpolation.}
    \label{fig:latticeBlocks}
\end{figure}

To exploit local coherence, we define $N_{\mathcal{A}}$ aggregates by partitioning the lattice sites into contiguous blocks, see Figure \ref{fig:latticeBlocks}\footnote{Throughout this work, we define an aggregate as a collection of lattice points.}. Inside each aggregate, we split the interpolation to account for the spin components, \begin{equation} \mathcal{A}_{i} \times \lbrace{0\rbrace}, \quad \mathcal{A}_{i}\times\lbrace1\rbrace.
\end{equation}The test vectors are then partitioned over the aggregates for each spin component and arranged in columns to build the interpolator, see the details in Algorithm \ref{alg:DDalpha}. This aggregation scheme implies that $P=\sigma_3 P \sigma_3$, which allows to preserve the $\sigma_3$-symmetry on the coarse grid operator, \begin{equation}
    D_c^\dagger = \sigma_3 D_c \sigma_3.
\end{equation}The lattice blocking for the aggregation does not need to coincide with the one for SAP. On the coarse grid, each site then possesses $2N_v$ degrees of freedom, \textit{i.e.}\ the coarse grid has a total of $2N_v N_{\mathcal{A}}$ variables. The quality of the interpolator $P$ is iteratively refined through $N_{b}$ bootstrap adaptivity iterations. We detail the two-level setup description in Algorithm \ref{alg:DDalpha}. \\

\begin{algorithm}[t]
\DontPrintSemicolon
\KwIn{$N_v$, $N_{b}$, $\nu$, aggregates $\{\mathcal{A}_j\}_{j=1}^{N_{\mathcal{A}}}$}
\KwOut{Interpolation operator $P$}

\textbf{Initialization:} Generate random test vectors $\{w_i\}_{i=1}^{N_v}$\;

\For{$i = 1,\dots,N_v$}{
    Apply $\nu$ SAP steps to approximately solve $Dx = 0$ with initial guess $w_i$\;
    Update $w_i \leftarrow x$\;
}

\tcp{Partition the test vectors over the aggregates}
\tcp{Loop over aggregates}
\For{$j=1,\dots,N_{\mathcal{A}}$}{
    \tcp{Loop over spin components}
    \For{$s=0,1$}{ 
        \tcp{Loop over test vectors}
        \For{$i=1,\dots,N_v$}{
        \tcp{k denotes a lattice site}
        $(w_i^{(j,s)})_{k} \leftarrow
        \begin{cases}
        (w_i)_{(k,s)}, & k \in \mathcal{A}_j,\\
        0, & \text{otherwise}
        \end{cases}$

        }
        Orthonormalize $\{w_i^{(j,s)}\}_{i=1}^{N_v}$\;
    }
    
}

\tcp{Construct the interpolation operator by arranging the partitioned vectors in columns}

$P \leftarrow \big( w_1^{(1,0)} \mid w_1^{(1,1)} \mid \cdots \mid w_{N_v}^{(1,0)} \mid w_{N_v}^{(1,1)} \mid \cdots \mid
w_1^{(N_{\mathcal{A}},0)} \mid w_1^{(N_{\mathcal{A}},1)} \mid \cdots \mid w_{N_v}^{(N_{\mathcal{A}},0)} \mid w_{N_v}^{(N_{\mathcal{A}},1)} \big)$\;

\tcp{Refine it with Nb bootstrap adaptivity iterations}
\For{$k=1,\dots, N_{b}$}{
    \For{$i=1,\dots,N_v$}{
        Apply one two-grid iteration with the current $P$ to approximately solve $Dx = (w_i - D w_i)$\;
        Update $w_i \leftarrow x$\;
    }
    Partition the test vectors over the aggregates\;
    Orthonormalize $\{w_i^{(j,s)}\}_{i=1}^{N_v}$\;
    Reconstruct $P$
}

\caption{DD$\alpha$AMG setup phase}
\label{alg:DDalpha}
\end{algorithm}

The aggregation scheme yields a coarse-grid operator $D_c$ that preserves the nearest-neighbor coupling of the Dirac operator. On the coarse grid, the operator takes the form \begin{equation}
    (D_c)_{xy} =\left[(m_0 +2)- A(x) \right]\delta_{xy} - \sum_{\mu=0}^1 \left(B_\mu(x) \delta_{x+\hat{\mu},y} + C_\mu(x) \delta_{x-\hat{\mu},y} \right),
\end{equation}where $x,y$ denote aggregates and $A(x),B_\mu(x), C_\mu(x)$ are the \textit{coarse gauge links}. At each lattice site, these links have $(2 N_v)^2$ components. These links are computed once during the setup phase and subsequently reused for every application of the coarse matrix, significantly reducing the overhead of the iterative solver. \\

To extend this framework to a multilevel hierarchy, the process is applied again. The current coarse lattice is aggregated and one generates a new set of test vectors by approximating $D_c x = 0$ at the current level. This continues until the grid is sufficiently small to be solved efficiently. In our case, we use GMRES without preconditioning at the coarsest level. Further algorithmic details and performance benchmarks, in the context of QCD, are in Ref.\ \cite{Frommer2014}. After several numerical experiments, we found that the adaptivity refinement does not improve the convergence properties in the Schwinger Model. Therefore, we worked only with smooth test vectors and set $N_b=0$. Once the interpolator and coarse-grid operator are built, DD$\alpha$AMG follows Algorithm \ref{alg:two}.

\section{Structured-Multiscale approach for Lattice QCD}\label{AM-AMG}
We now look at a different way to solve eq.\  \eqref{eq:DiracEq}. The method is called Structured-Multiscale Algebraic Multigrid \cite{ehrmannAMAMG}, which is a setup method for AMG and follows the idea of Algebraic Multiscale (AMS) for reservoir modeling \cite{zhou}. AMS uses the cells from a given reservoir model and groups them together to so-called wirebaskets to have smaller coarse grids. Other ways of constructing an AMS prolongation are, \textit{e.g.}\ using prolongator smoothing, which can be applied to a purely algebraic, simple initial prolongator \cite{MOYNER201646, silvia153}.  More literature connected to AMS methods with different ideas of constructing the wirebaskets, using geometric coarsening, can be found in \cite{lie2016successful, 10.2118/182694-MS, moyner2015multiscale,MOYNER201646,ene2015,ene2016,wang2014algebraic}.

\subsection{SM-AMG for LQCD}
SM-AMG uses lattice geometry to construct compact aggregates with multiple function types. Overlaps of the aggregate borders improve interpolation accuracy per aggregate and by using the different function types yield fine-grained control of coarsening.  \\

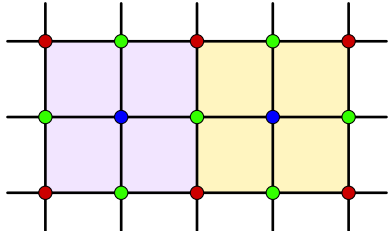
\begin{figure}[h]
    \centering
    \definecolor{qqqqff}{rgb}{0,0,1}
\definecolor{ffdxqq}{rgb}{1,0.8431372549019608,0}
\definecolor{cczzff}{rgb}{0.8,0.6,1}
\definecolor{ttffqq}{rgb}{0.2,1,0}
\definecolor{ccqqqq}{rgb}{0.8,0,0}
\begin{tikzpicture}[line cap=round,line join=round,>=triangle 45,x=1cm,y=1cm]
\clip(-1,-0.5) rectangle (5,3);
\fill[line width=0.8pt,color=cczzff,fill=cczzff,fill opacity=0.25] (0,0) -- (0,1) -- (0,2) -- (1,2) -- (2,2) -- (2,1) -- (2,0) -- (1,0) -- cycle;
\fill[line width=0.8pt,color=ffdxqq,fill=ffdxqq,fill opacity=0.25] (2,0) -- (3,0) -- (4,0) -- (4,1) -- (4,2) -- (3,2) -- (2,2) -- (2,1) -- cycle;
\draw [line width=1pt] (0,1)-- (1,1);
\draw [line width=1pt] (2,1)-- (1,1);
\draw [line width=1pt] (1,1)-- (1,2);
\draw [line width=1pt] (1,1)-- (1,0);
\draw [line width=1pt] (3,0)-- (3,1);
\draw [line width=1pt] (3,1)-- (4,1);
\draw [line width=1pt] (3,1)-- (3,2);
\draw [line width=1pt] (3,1)-- (2,1);
\draw [line width=1pt] (0,1)-- (-0.5,1);
\draw [line width=1pt] (0,2)-- (-0.5,2);
\draw [line width=1pt] (0,0)-- (-0.5,0);
\draw [line width=1pt] (0,0)-- (0,-0.5);
\draw [line width=1pt] (1,0)-- (1,-0.5);
\draw [line width=1pt] (2,0)-- (2,-0.5);
\draw [line width=1pt] (3,0)-- (3,-0.5);
\draw [line width=1pt] (4,0)-- (4,-0.5);
\draw [line width=1pt] (4,0)-- (4.5,0);
\draw [line width=1pt] (4,1)-- (4.5,1);
\draw [line width=1pt] (4,2)-- (4.5,2);
\draw [line width=1pt] (4,2)-- (4,2.5);
\draw [line width=1pt] (3,2)-- (3,2.5);
\draw [line width=1pt] (2,2)-- (2,2.5);
\draw [line width=1pt] (1,2)-- (1,2.5);
\draw [line width=1pt] (0,2)-- (0,2.5);
\draw [line width=1pt] (0,0)-- (0,1);
\draw [line width=1pt] (0,1)-- (0,2);
\draw [line width=1pt] (0,2)-- (1,2);
\draw [line width=1pt] (1,2)-- (2,2);
\draw [line width=1pt] (2,2)-- (2,1);
\draw [line width=1pt] (2,1)-- (2,0);
\draw [line width=1pt] (2,0)-- (1,0);
\draw [line width=1pt] (1,0)-- (0,0);
\draw [line width=1pt] (2,0)-- (3,0);
\draw [line width=1pt] (3,0)-- (4,0);
\draw [line width=1pt] (4,0)-- (4,1);
\draw [line width=1pt] (4,1)-- (4,2);
\draw [line width=1pt] (4,2)-- (3,2);
\draw [line width=1pt] (3,2)-- (2,2);
\begin{scriptsize}
\draw [fill=ccqqqq] (0,0) circle (2.5pt);
\draw [fill=ttffqq] (0,1) circle (2.5pt);
\draw [fill=ccqqqq] (0,2) circle (2.5pt);
\draw [fill=ttffqq] (1,2) circle (2.5pt);
\draw [fill=ccqqqq] (2,2) circle (2.5pt);
\draw [fill=ttffqq] (2,1) circle (2.5pt);
\draw [fill=ccqqqq] (2,0) circle (2.5pt);
\draw [fill=ttffqq] (1,0) circle (2.5pt);
\draw [fill=ttffqq] (3,0) circle (2.5pt);
\draw [fill=ccqqqq] (4,0) circle (2.5pt);
\draw [fill=ttffqq] (4,1) circle (2.5pt);
\draw [fill=ccqqqq] (4,2) circle (2.5pt);
\draw [fill=ttffqq] (3,2) circle (2.5pt);
\draw [fill=qqqqff] (1,1) circle (2.5pt);
\draw [fill=qqqqff] (3,1) circle (2.5pt);
\end{scriptsize}
\end{tikzpicture}
    \caption[Example of wirebaskets with four vertices, four edges and one interior point on a uniformly structured grid]{Example of wirebaskets (purple and yellow) with four vertices (red), four edges (green) and one interior point (blue) on a uniformly structured grid.}
    \label{fig:wirebasket}
\end{figure}

To coarsen, the grid is divided into sets, which are called wirebaskets. Wirebaskets are aggregates that are distinguished by additional classification of the grid points into vertices $V$, edges $E$ and interior points $I$. Figure \ref{fig:wirebasket} shows an example of wirebaskets with four vertices, four edges and one interior point marked in red, green and blue, respectively. In this case, we have the minimal wirebasket size of nine grid points. All classifications have a different ``role", where the vertices will transfer to the coarse grid. The edges are directly interpolated from the vertices and only have couplings to vertices and/or to other neighboring wirebaskets, while the interior points only have couplings inside one wirebasket.\\

These function types are used for the setup of the local interpolation operator in each aggregate. In an ideal scenario, the interpolation would be equivalent to the Schur-complement between coarse and fine variables. It is possible and necessary in SM-AMG to reuse variables in multiple aggregates, which is a strength of SM-AMG combined with the non-constant interpolation in each aggregate.\\

To compute the interpolation operator, we follow the idea of AMS \cite{zhou} and by that reformulate the linear system (\ref{eq:problem1}) block-wise for one aggregate with the different 
function types. The rewritten form
\begin{align}
    \begin{pmatrix}
    A_{II} & A_{IE} & A_{IV} \\
    A_{EI} & A_{EE} & A_{EV}\\
    A_{VI} & A_{VE} & A_{VV}
    \end{pmatrix} 
    \begin{pmatrix}
    u_{I} \\ u_{E} \\ u_{V}
    \end{pmatrix}
    \,=\,
    \begin{pmatrix}
    f_{I} \\f_{E}\\ f_{V}
    \end{pmatrix} \label{eq:AMAMGbase}
\end{align}
does not apply any simplifications based on the application field, grid structure or physical information. We want to solve (\ref{eq:AMAMGbase}) based on the wirebaskets and the two-level scheme described in Section \ref{AMG}. We assume ideal per aggregate pre-smoothing of the solution of equation (\ref{eq:AMAMGbase}), which leads 
to a zero residual for the interior and edge variables. \\

The coarse level defect solution vector $u_{V}^{H}$ corresponds to the solution of the vertex variables $u_{V}$, which helps us reformulate the system. We do not have to consider the couplings from the vertices to interior points $A_{VI}$ and from the vertices to the edges $A_{VE}$, since the vertices are coarse grid variables and are interpolated by identity. This gives us the reformulated system
\begin{align}
    \begin{pmatrix}
        A_{II} & A_{IE} & A_{IV} \\
        A_{EI} & A_{EE} & A_{EV}\\
        0 & 0 & I_{VV}
    \end{pmatrix}
    \begin{pmatrix}
        u_{I} \\ u_{E} \\ u_{V}
    \end{pmatrix}
    \,=\, 
    \begin{pmatrix}
        0 \\ 0 \\ u_{V}^{H}
    \end{pmatrix}. \label{eq:AMAMG2}
\end{align}
Now, we solve the first row of (\ref{eq:AMAMG2}) for $u_{I}$ and get that $u_{I}\,=\,-A_{II}^{-1}A_{IV}u_{V}^{H}-A_{II}^{-1}A_{IE}u_{E}$ holds. We then combine this formulation of $u_{I}$ with the second row of (\ref{eq:AMAMG2}) to get 
\begin{align*}
 u_{E}\,&=\,-\hat{A}_{EE}^{-1}(A_{EV}-A_{EI}A_{II}^{-1}A_{IV})u_{V}^{H}, \nonumber
\end{align*}
with $\hat{A}_{EE}=A_{EE}-A_{EI}A_{II}^{-1}A_{IE}$. This yields a fully algebraic formulation for the interpolation of edges, based only on the coarse level solution for the vertices and the couplings inside the aggregate.\\

We find the formula for the interior points in the same way and therefore solve the second row of (\ref{eq:AMAMG2}) for $u_{E}$ and get $u_{E}\,=\,-A_{EE}^{-1}A_{EI}u_{I}-A_{EE}^{-1}A_{EV}u_{v}^{H}$,
which is inserted in the first row and solved for $u_{I}$. This yields
\begin{align*}
 u_{I}\,&=\,-\hat{A}_{II}^{-1}(A_{IV}-A_{IE}A_{EE}^{-1}A_{EV})u_{V}^{H}, \nonumber
\end{align*}
with $\hat{A}_{II}=A_{II}-A_{IE}A_{EE}^{-1}A_{EI}$.\\ 

Combining the formula of $u_{I}$ and $u_{E}$ gives the interpolation operator for SM-AMG, \textit{i.e.}\
\begin{align}
    \mathcal{P}_{\textrm{SM-AMG}}\,=\,\mathcal{G}\mathcal{N}\cdot
    \begin{pmatrix}
        -\hat{A}_{II}^{-1}(A_{IV}-A_{IE}A_{EE}^{-1}A_{EV})\\
        -\hat{A}_{EE}^{-1}(A_{EV}-A_{EI}A_{II}^{-1}A_{IV})\\
        I_{VV}
    \end{pmatrix} \label{eq:AMAMGP}
\end{align}
holds. $\mathcal{G}$ describes the reordering of (\ref{eq:problem1}) in the function types of (\ref{eq:AMAMGbase}), which is never actually used in the implementation, but helps for readability. $\mathcal{N}$ is the normalization of interpolation formulas such that a constant vector would be interpolated as a constant vector.\\

The interpolation is then local for each aggregate. We only have conflicts with the edges, since they can be in more than one aggregate. To prevent this, we assign the edges to the first aggregate they were assigned to as the one primary aggregate. From this we interpolate the edge and in all other aggregates we do not set up an interpolation formula. One might think that the vertices may cause some conflicts as well, since they also belong to more than one aggregate. This is not the case, since the vertices span the coarse level and therefore, the interpolation formula for the vertices is the identity.\\

The global interpolator is built using the local interpolation operator (\ref{eq:AMAMGP}) and is then used in a V-cycle as described in Algorithm \ref{alg:two}.

\subsection{Current Implementation}\label{subsec:AMAMG}
Currently, this method is implemented in the fast solver software SAMG (Fraunhofer SCAI). SAMG is a software package for algebraic multigrid and its applications. It has a wide range of smoothers, preconditioners, coarsening schemes and interpolation schemes to choose from. We used this to try different combinations to obtain the best results.\\

This software is not meant for QCD and its usage has fundamental differences with QCD codes. For instance, SAMG requires an explicit matrix to be assembled in CSR format, which is generally unfeasible in Lattice Field Theory. However, as we mentioned before, we  test SM-AMG as a proof of concept for the Schwinger Model with an already working, tested and reliable software.\\

In SAMG, the method is implemented with an algebraic approach. This means that the coarsening is based on the matrix adjacency graph, not the underlying geometry of the lattices. The precise way on how to coarsen algebraically is called Aggregative-Multiscale AMG (AM-AMG) and is described in detail by Gries \cite{dissAMAMG}, Chapter 3.\\

\begin{figure}[h]
    \centering
    \begin{subfigure}[t]{0.4\textwidth}
        \centering
        \includegraphics[width=1\linewidth]{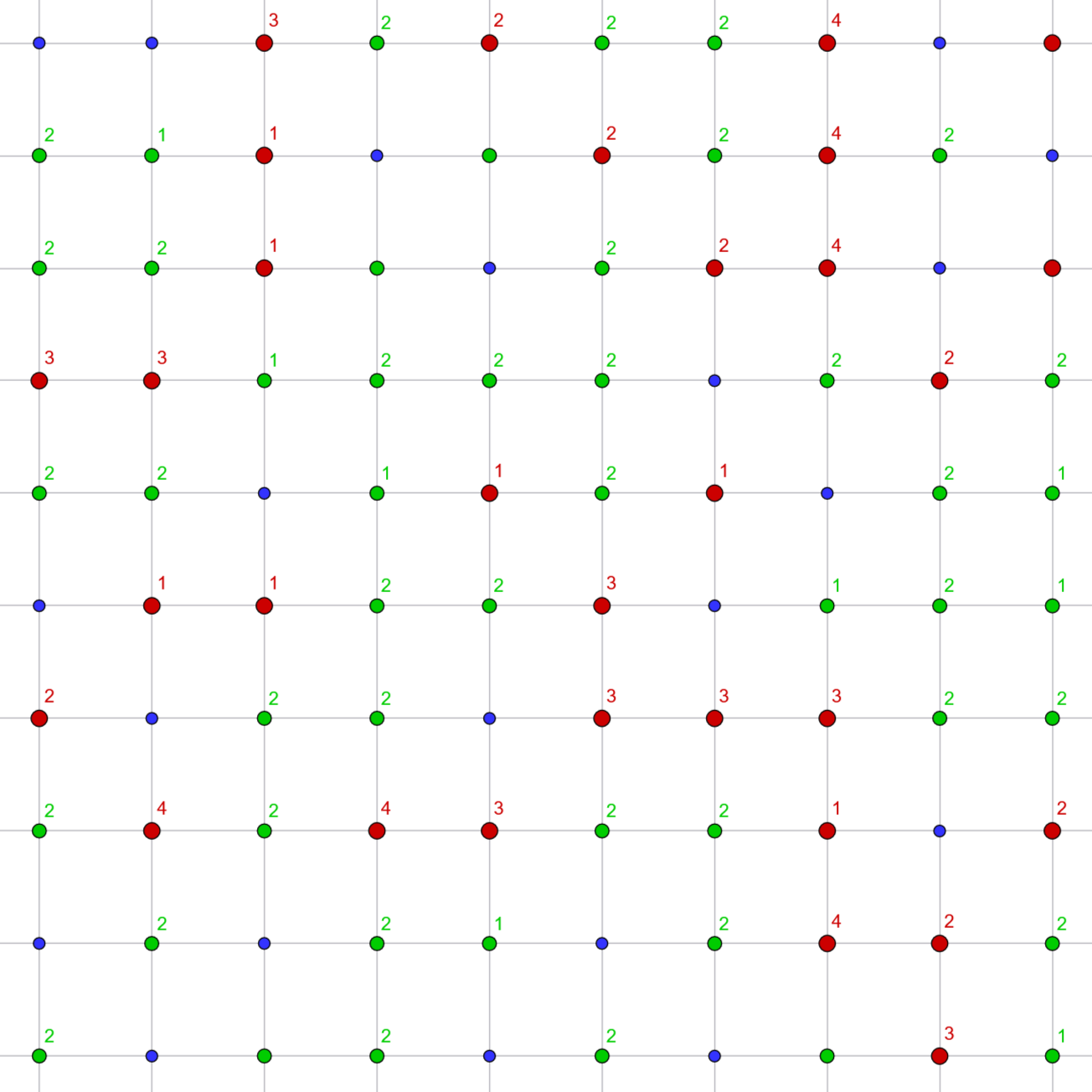}
    \end{subfigure}%
    ~ 
    \begin{subfigure}[t]{0.4\textwidth}
        \centering
        \includegraphics[width=1\linewidth]{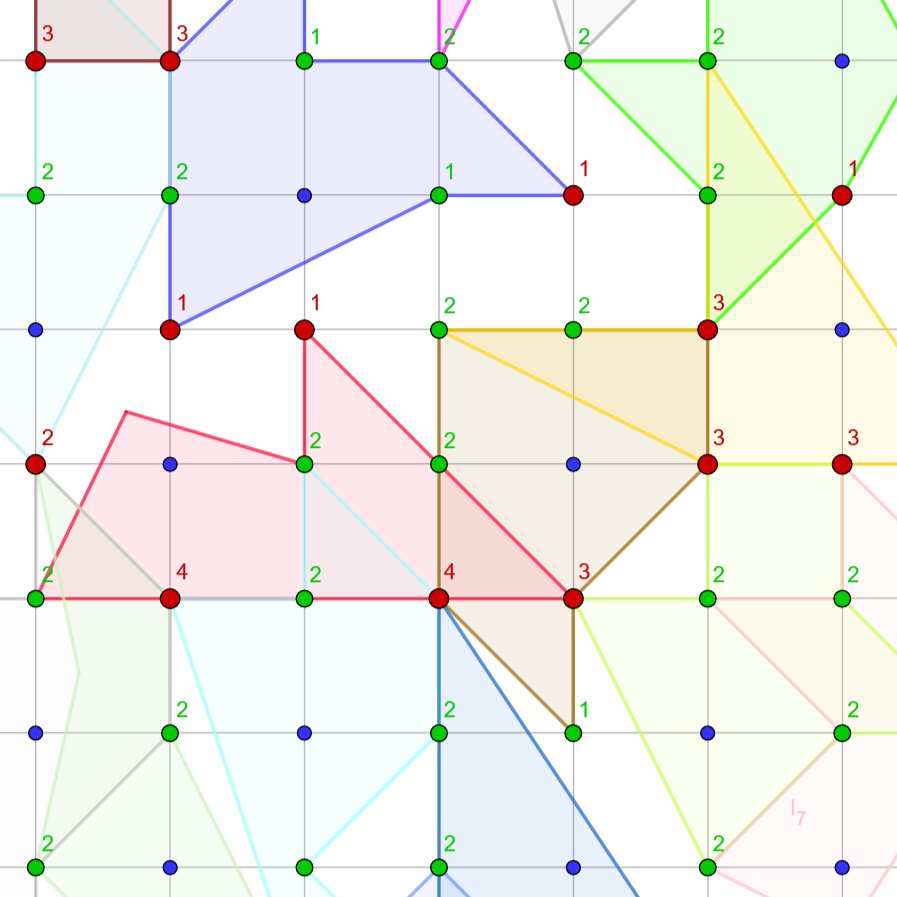}
    \end{subfigure}
    \caption{Example of wirebaskets for the Schwinger model with $\beta=4$ and $V=64^2$. The colors of the points mark the function types in red, green and blue as vertices, interior points and edges, respectively. On the left, there is a cutout of size $10^{2}$ with the number denoting the number of wirebaskets the respective point is in. On the right, there is a close up of some wirebaskets colored in, where each color corresponds to one wirebasket.  }
    \label{fig:wirebaskets_scatter}
\end{figure}

The Schwinger model and the Dirac equation have an underlying geometry that we are not exploiting so far. We can see this by looking at the algebraically built wirebaskets in Figure \ref{fig:wirebaskets_scatter}. In the figure on the right, each colored in area belongs to one wirebasket, while points can belong to more than one wirebasket following the rule of the function types. Note that this is purely algebraic and does not mean that any area physically belongs to only one aggregate. The colored in areas only illustrate the connection between points in an aggregate. We observe that the wirebaskets are not compactly shaped rectangles as we would have expected looking at Figure \ref{fig:wirebasket}. Instead, the wirebaskets are shaped in an algebraic way, while taking the neighbors into account, but clearly not the geometry. This is a disadvantage for the convergence of the method, since the geometry provides information that is lost in algebraic coarsening. Therefore, taking geometry into account is a future step for improving the convergence of AM-AMG.\\ 

Since SAMG is a real solver, the AM-AMG implementation is also real valued. To adapt this to Lattice QCD, in our case the Schwinger model, we used the K-formulation \cite{day2001solving} to rewrite the complex matrix as a real valued matrix. For the K-formulation of a complex valued matrix, we rewrite each complex entry $c_{ij}=a_{ij}+ib_{ij}$ as a $2\times 2$ real valued block of the form 
\begin{align*}
    \begin{pmatrix}
                a_{ij} & -b_{ij} \\ b_{ij} & a_{ij}
            \end{pmatrix}.
\end{align*}  
Therefore, in our implementation the used matrix is always twice as big as the given problem size, which can lead to a huge amount of memory needed for big problem sizes. Memory bandwidth also differs considerably from that in DD$\alpha$AMG. For these reasons, when comparing both methods we only look at the FLOP count and not time to solution. 
\\

With this, we have to keep in mind that the evaluation of AM-AMG is just a first impression and a starting point for further development. However, the first results in the following show that AM-AMG could be a step in the right direction. \\

\section{Numerical Results}\label{results}
We implemented an MPI-parallel HMC to simulate the two-flavor Schwinger model at different $\beta=1/g^2$ (not to be confused with the spin index)\footnote{We work in lattice units, \textit{i.e.}\ we fix the lattice spacing $a=1$ and control the continuum limit with $\beta\rightarrow \infty$. In general $\beta=1/(a^2g^2)$, where $g$ is the gauge coupling.}, and for lattices of size $128^2,256^2,512^2$ and $1024^2$. We use the bare masses in Table \ref{tab::CriticalValues} as a reference for creating gauge configurations close to the chiral limit. For each volume, we simulated the Schwinger model at $\beta=2, 4$ and 6 for three different masses, roughly at $|m_0-m_c|\sim0.1,0.01$ and $0.001$. For each case we generated 10 gauge configurations. \\

\begin{table}[h]
\centering
\begin{tabular}{@{}cc@{}}
\toprule
$\beta$ & $m_{\textrm{c}}$ \\ \midrule
2                & -0.1968(9)            \\
4                & -0.1033(1)            \\
6                & -0.0719(1)            \\ \bottomrule
\end{tabular}
\caption{Critical values of $m_0$ for the Schwinger model \cite{Christian2006}.}
\label{tab::CriticalValues}
\end{table}

\begin{figure}[h]
    \centering
    \begin{subfigure}[t]{0.5\textwidth}
        \centering
        \includegraphics[width=1\linewidth]{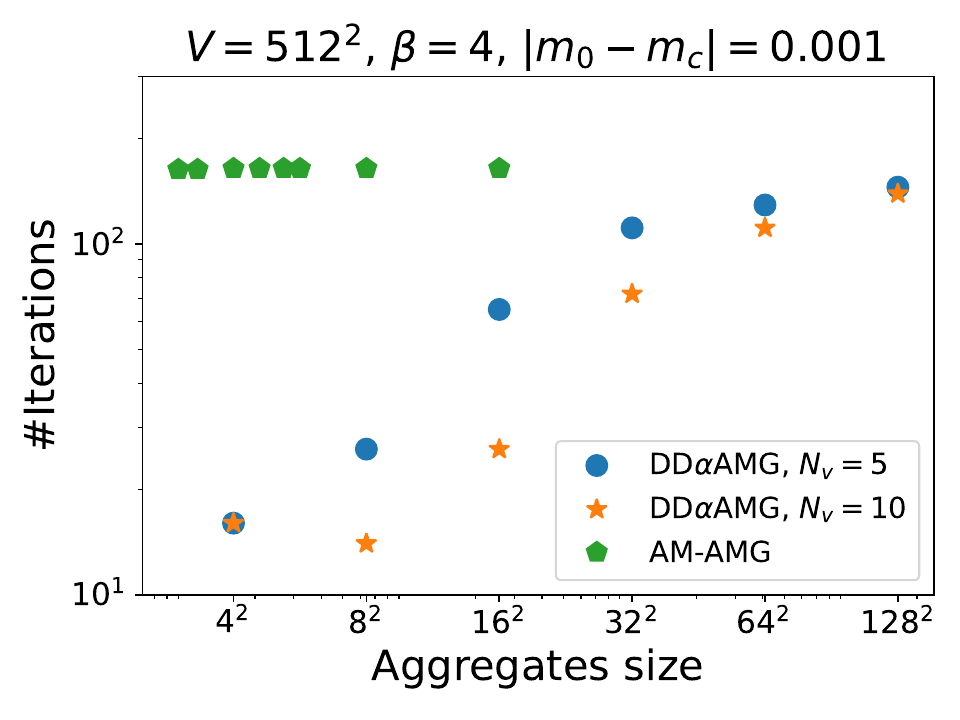}
    \end{subfigure}%
    ~ 
    \begin{subfigure}[t]{0.5\textwidth}
        \centering
        \includegraphics[width=1\linewidth]{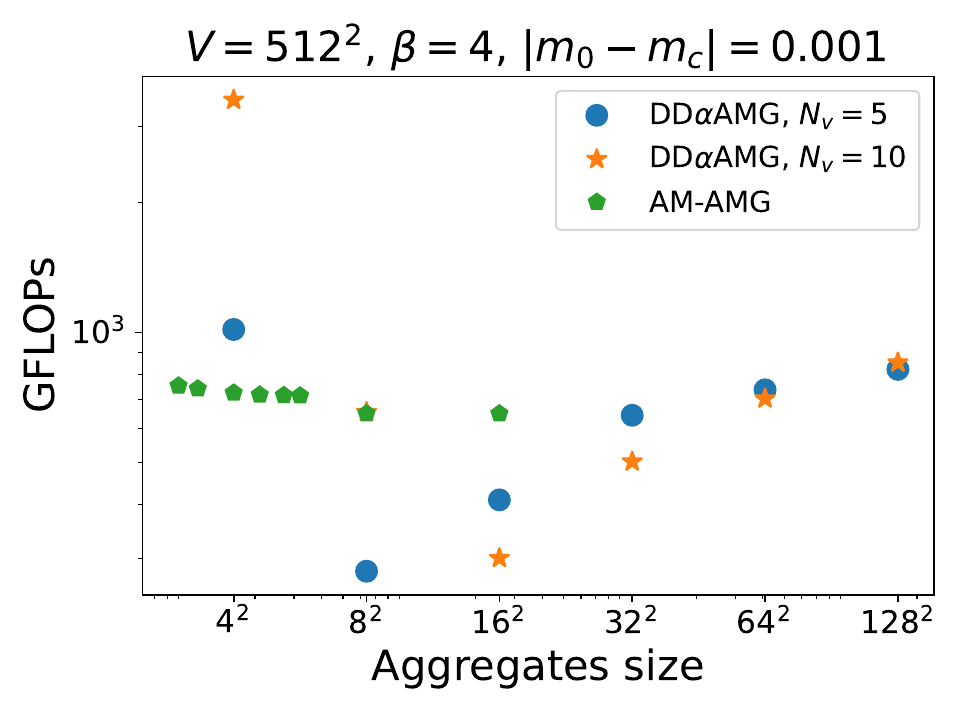}
    \end{subfigure}
    \caption{We show the variation of the fine-grid iterations (left) and the FLOP count (right) when changing the parameters. The size of the aggregates refers to wirebaskets for AM-AMG and to lattice blocks for DD$\alpha$AMG. For the latter we also vary the number of test vectors. Both methods were used as preconditioners. The results are for only one configuration and one right-hand side at $\beta=4$, $V=512^2$ and $|m_0-m_c|=0.001$, since the behavior is generic. For a small block size and large $N_v$, the interpolator of DD$\alpha$AMG captures the near-kernel very well. However, the setup phase and solving the problem on the coarse grid become more expensive. The parameters have to be fine-tuned in order to find the best performance. For AM-AMG the solver is not sensitive to the size of the wirebaskets. The FLOP count includes the setup phase.}
    \label{fig:parameters_spread}
\end{figure}

We test two-level methods. The parameter values for DD$\alpha$AMG are listed in Table \ref{tab::DDalphaParameters}. As explained in Section \ref{subsec:AMAMG}, instead of the structured approach the algebraic approach AM-AMG is evaluated in the numerical simulations. For AM-AMG each wirebasket has 9 elements and are built algebraically. We employ the same set of parameters for each test case. In practice, the convergence behavior of DD$\alpha$AMG is strongly affected by these settings, and a careful fine‑tuning of the parameters is usually required to achieve the best performance. This is not the case with AM-AMG, where convergence does not depend strongly on the size of the wirebaskets. We further discuss this issue in Figure \ref{fig:parameters_spread}, where we explicitly show the effect of the parameters on each solver. To make DD$\alpha$AMG and AM-AMG as comparable as possible, we use them as preconditioners for FGMRES and fix the relative convergence tolerance to $10^{-10}$. Both preconditioners use a block solver as a smoother with 2 post-smoothing steps, GMRES as the coarsest-level solver with a relative tolerance of $10^{-1}$ and a V-cycling strategy.\\
\begin{table}[h]
\centering
\begin{tabular}{@{}ccc@{}}
\toprule
      & Size of lattice blocks & $8^2$     \\
      & Number of test vectors                & 10        \\
$l=1$ & Pre-smoothing steps                   & 0         \\
      & Post-smoothing steps                  & 2         \\
      & SAP block solves tolerance            & $10^{-3}$ \\ \midrule
      & Number of GMRES restarts              & 10        \\
$l=2$ & Restart length                        & 400       \\
      & Relative residual tolerance           & 0.1       \\ \bottomrule
\end{tabular}
\caption{Parameters for DD$\alpha$AMG. For each combination $(\beta,V,m_0)$ we use the exact same parameters.}
\label{tab::DDalphaParameters}
\end{table}

\begin{figure}[h]
    \centering
    \begin{subfigure}[t]{0.5\textwidth}
        \centering
        \includegraphics[width=1\linewidth]{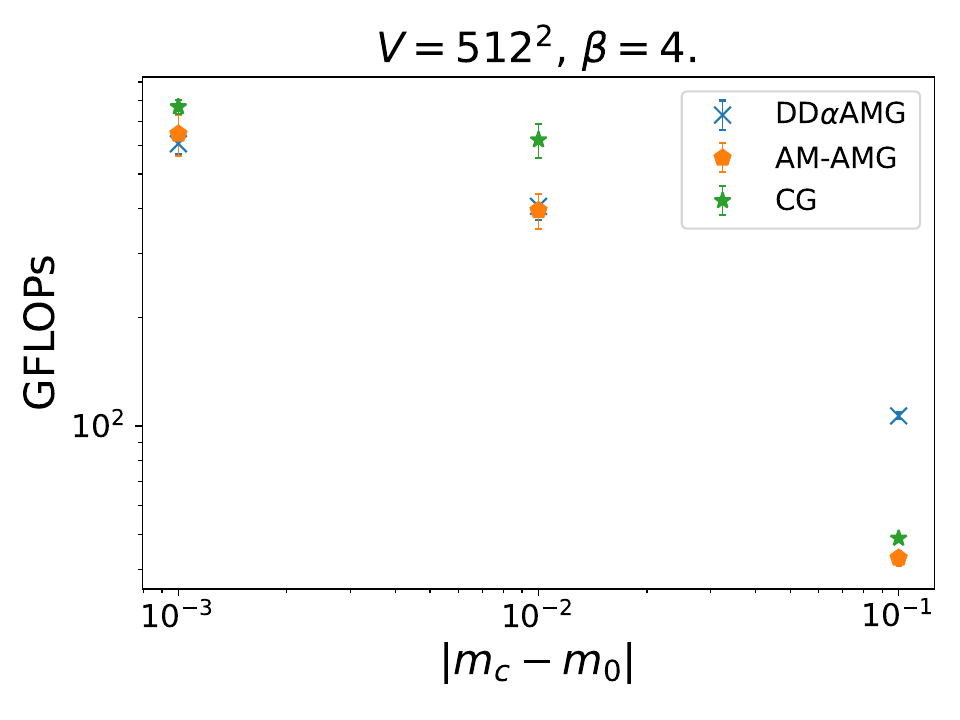}
    \end{subfigure}%
    ~ 
    \begin{subfigure}[t]{0.5\textwidth}
        \centering
        \includegraphics[width=1\linewidth]{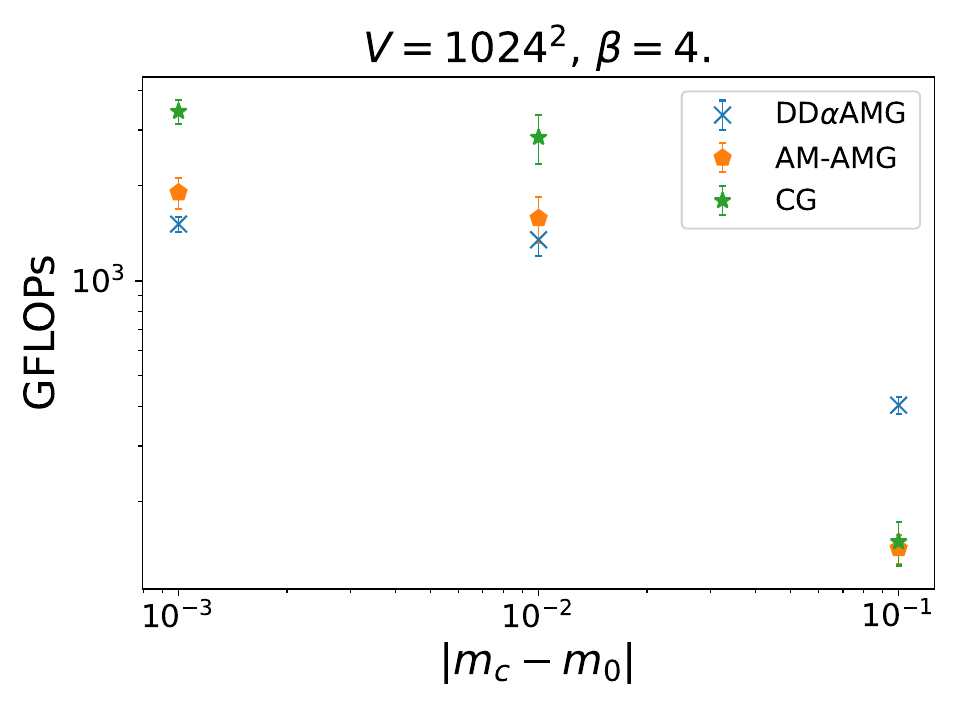}
    \end{subfigure}
    \caption{A comparison of the floating-point operations it takes to achieve convergence for both preconditioners as a function of the mass shift. The results are averaged over 10 decorrelated configurations and 10 different random right-hand sides. For DD$\alpha$AMG we use the parameters of Table \ref{tab::DDalphaParameters} and for AM-AMG the size of the wirebaskets is set to 9.}
    \label{fig:b4_plot}
\end{figure}

We solve the associated matrix problem for each configuration using ten randomly generated right-hand sides. We then average the iteration number and FLOP count over the configurations and the right-hand sides for each combination $(\beta,V,m_0)$. The resulting means together with their deviations are listed in Tables \ref{tab::TabIterations} and \ref{tab::TabFlops} for $\beta = 2,4$ and $6$. For every configuration and right-hand side, we repeat the setup phase and consider its cost for the FLOP count. We also present results obtained with the Conjugate Gradient (CG) solver applied to the normal equation $D^\dagger Dx = D^\dagger\psi$; this provides a quantitative measure of the ill‑conditioning of the system. In Figure \ref{fig:b4_plot}, we show the GFLOP count for large volumes at $\beta=4$. Data for other examples are given in Table \ref{tab::TabFlops}.\\

\begin{figure}[h]
    \centering
    \begin{subfigure}[t]{0.5\textwidth}
        \centering
        \includegraphics[width=1\linewidth]{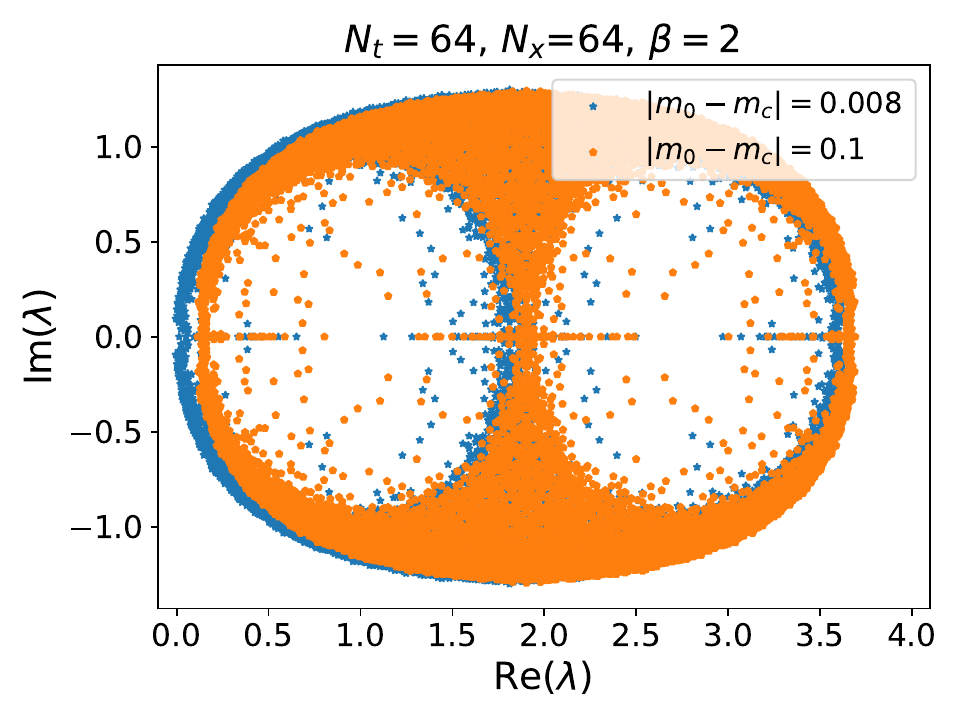}
        \caption{Eigenvalue spectrum for a coarse lattice.}
    \end{subfigure}%
    ~ 
    \begin{subfigure}[t]{0.5\textwidth}
        \centering
        \includegraphics[width=1\linewidth]{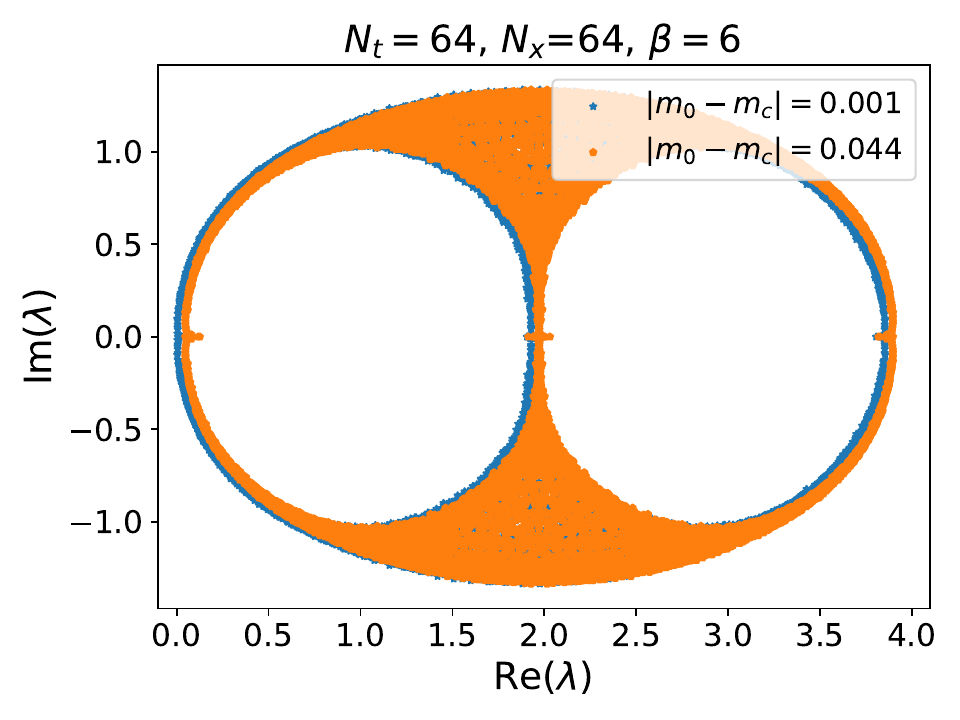}
        \caption{Eigenvalue spectrum for a fine lattice.}
    \end{subfigure}
    \caption{Eigenvalue spectra for the Dirac matrix associated with configurations generated at strong and weak coupling on a lattice of size $64^2$. For small $\beta$, the spectrum is more spread and close to the critical mass the density of small eigenvalues increases drastically.}
    \label{fig:eigenvalues}
\end{figure}

\begin{figure}[h]
    \centering
    \begin{subfigure}[t]{0.5\textwidth}
        \centering
        \includegraphics[width=1\linewidth]{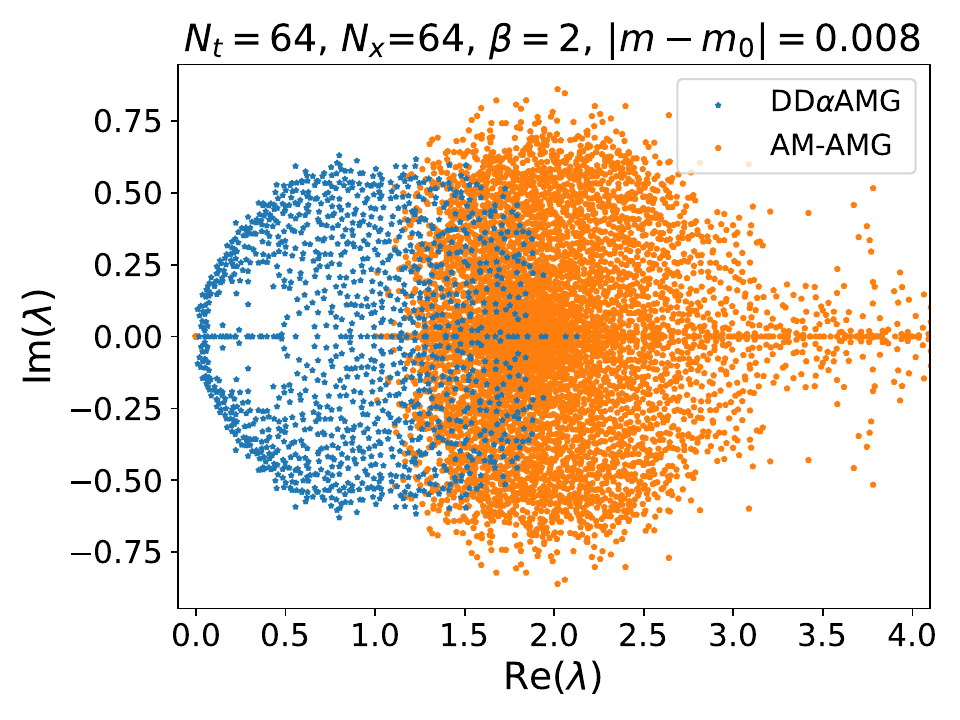}
    \end{subfigure}%
    ~ 
    \begin{subfigure}[t]{0.5\textwidth}
        \centering
        \includegraphics[width=1\linewidth]{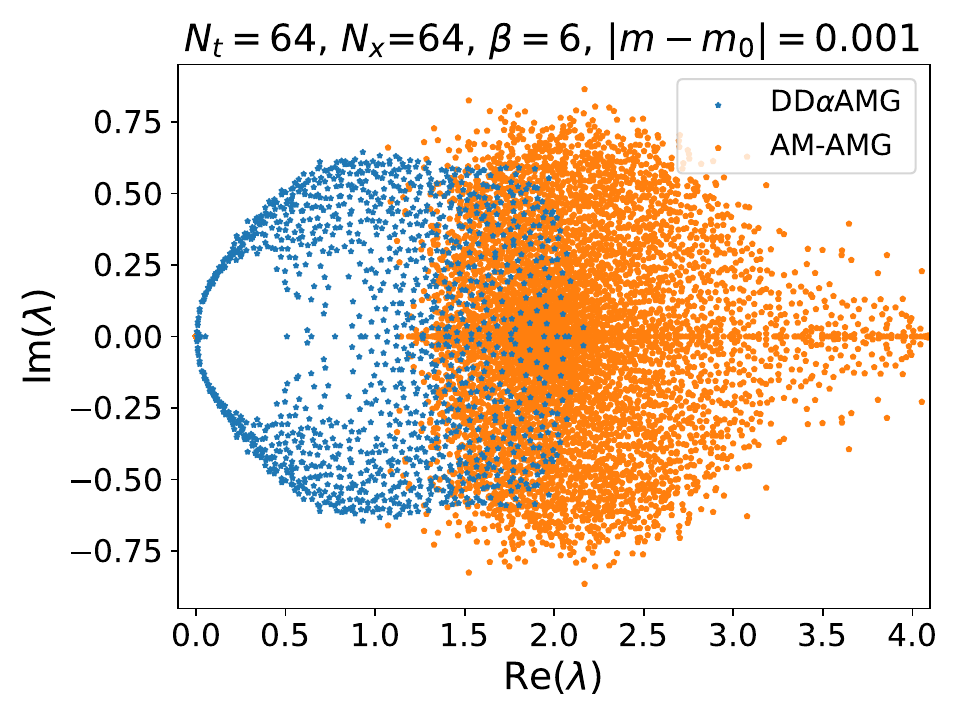}
    \end{subfigure}
    \caption{Eigenvalue spectra for the coarse-grid matrices associated with configurations generated at strong and weak coupling on a lattice of size $64^2$. We compare the coarse-grid operator of both methods when applied to the same gauge configuration.}
    \label{fig:eigenvaluesDc}
\end{figure}

In Figure \ref{fig:eigenvalues} we show the eigenvalue spectrum of $D$ for small and large $\beta$. For strong coupling (small $\beta$), the spectrum spreading becomes more pronounced. This influences the Dirac matrix conditioning and the algorithms efficiency, since the density of small eigenvalues increases for coarse lattices. In Figure \ref{fig:eigenvaluesDc} we show the spectrum of the coarse-grid matrix associated to each multilevel method. \\

For $\beta=4$ and $6$, Tables \ref{tab::Beta4Flops} and \ref{tab::Beta6Flops}, we observe that the two-level DD$\alpha$AMG and AM-AMG have a comparable number of operations and both involve less computational effort than CG close to the critical mass, particularly for the large volumes. The number of FGMRES iterations is very stable for DD$\alpha$AMG in all cases (see Tables \ref{tab::Beta4It} and \ref{tab::Beta6It}), indicating that the low-modes are well represented on the coarse grid, where the problem becomes more challenging to solve the closer we move to the critical mass. As a consequence, most of the operations are performed on the coarsest level. On the other hand, AM-AMG has a larger and slightly fluctuating number of iterations on the fine grid, which implies the method does not capture the near-kernel as well as DD$\alpha$AMG does. However, although the number of iterations required by AM-AMG exceeds that of DD$\alpha$AMG, the overall computational cost, measured in floating-point operations, remains comparable. Furthermore, the iteration count exhibits only a weak dependence on the lattice volume when $m_{0}$ is held fixed. The setup phase of AM-AMG is significantly less expensive, and the algorithm requires minimal tuning: only the size of the wirebaskets must be specified.  As illustrated in Figure \ref{fig:parameters_spread}, the performance is essentially insensitive to this choice.  \\

For the coarse lattice at $\beta=2$, the DD$\alpha$AMG preconditioner yields once again a stable and low number of iterations on the fine grid, as demonstrated in Table \ref{tab::Beta2It}. Consequently, analogous to the behavior observed for other values of $\beta$, the majority of the computational effort is concentrated on the coarse grid. For the large volumes, the performance would benefit from a third level with a proper tuning of the aggregates size. The AM-AMG preconditioner has a different behavior in this case. Both the fine-grid iteration count and the number of operations become substantially larger and display significant variability in the critical regime.  This indicates that the purely algebraic approach of AM-AMG, which does not rely on any physical information, becomes problematic for small masses or requires fine-tuning. Nevertheless, these observations represent only an initial exploration of the method, with the observed difficulties arising in the most challenging scenarios: A coarse lattice spacing combined with a bare mass close to the chiral limit.

\begin{table*}[t]
\begin{subtable}{\textwidth}
\centering
\begin{tabular}{@{}ccccc@{}}
\toprule
$V$      & $|m_0-m_c|$ & CG            & DD$\alpha$AMG & AM-AMG      \\ \midrule
$128^2$  & $0.0084$    & $3665\pm337$  & $15$      & $176\pm114$ \\
$256^2$  & $0.0084$    & $5641\pm583$  & $16$      & $190\pm60$  \\
$512^2$  & $0.0084$    & $8725\pm1521$ & $15$      & $383\pm162$ \\\midrule
$128^2$  & $0.01$   
& $3358\pm299$  & $14$      & $88\pm12$   \\
$256^2$  & $0.01$      & $6505\pm677$  & $16$      & $870\pm306$ \\
$512^2$  & $0.01$      & $7856\pm804$  & $15$      & $607\pm847$ \\
$1024^2$ & $0.01$      & $10486\pm776$ & $18$      & $559\pm536$ \\ \midrule
$128^2$  & $0.1$       & $568\pm34$    & $8$       & $10$    \\
$256^2$  & $0.1$       & $634\pm44$    & $9$       & $10$    \\
$512^2$  & $0.1$       & $639\pm38$    & $8$       & $10$    \\
$1024^2$ & $0.1$       & $715\pm37$    & $9$       & $10$    \\ \bottomrule
\end{tabular}
\caption{$\beta=2$}
\label{tab::Beta2It}
\end{subtable}

\vspace{0.5 mm}

\begin{subtable}{\textwidth}
\centering
\begin{tabular}{@{}ccccc@{}}
\toprule
$V$      & $|m_0-m_c|$ & CG            & DD$\alpha$AMG & AM-AMG     \\ \midrule
$128^2$  & $0.001$     & $3942\pm232$  & $14$      & $140\pm19$ \\
$256^2$  & $0.001$     & $7182\pm888$  & $16$      & $143\pm29$ \\
$512^2$  & $0.001$     & $7244\pm318$  & $15$      & $140\pm17$ \\
$1024^2$ & $0.001$     & $8099\pm697$  & $16$      & $103\pm11$ \\ \midrule
$128^2$  & $0.01$      & $2819\pm246$  & $13$      & $65\pm5$   \\
$256^2$  & $0.01$      & $4724\pm531$  & $14$      & $93\pm9$   \\
$512^2$  & $0.01$      & $5866\pm638$  & $14$      & $86\pm9$   \\
$1024^2$ & $0.01$      & $6703\pm1170$ & $15$      & $85\pm14$  \\ \midrule
$128^2$  & $0.1$       & $456\pm15$    & $8$       & $9$    \\
$256^2$  & $0.1$       & $481\pm11$    & $8$       & $9$    \\
$512^2$  & $0.1$       & $459\pm7$     & $7$       & $8$    \\
$1024^2$ & $0.1$       & $351\pm55$    & $7$       & $7$    \\ \bottomrule
\end{tabular}
\caption{$\beta=4$}
\label{tab::Beta4It}
\end{subtable}

\begin{subtable}{\textwidth}
\centering
\begin{tabular}{@{}ccccc@{}}
\toprule
$V$      & $|m_0-m_c|$ & CG             & DD$\alpha$AMG & AM-AMG     \\ \midrule
$128^2$  & $0.001$     & $4490\pm186$   & $14$      & $134\pm16$ \\
$256^2$  & $0.001$     & $6849\pm691$   & $15$      & $128\pm12$ \\
$512^2$  & $0.001$     & $7884\pm286$   & $15$      & $128\pm9$  \\
$1024^2$ & $0.001$     & $10868\pm1544$ & $16$      & $138\pm22$ \\ \midrule
$128^2$  & $0.01$      & $3170\pm188$   & $13$      & $64\pm4$   \\
$256^2$  & $0.01$      & $5120\pm722$   & $14$      & $83\pm9$   \\
$512^2$  & $0.01$      & $4935\pm551$   & $14$      & $67\pm4$   \\
$1024^2$ & $0.01$      & $5994\pm647$   & $14$      & $70\pm7$   \\ \midrule
$128^2$  & $0.0438$    & $950\pm22$     & $11$      & $18$   \\
$256^2$  & $0.0438$    & $1035\pm40$    & $11$      & $17$   \\
$512^2$  & $0.0438$    & $690\pm128$    & $8$       & $11\pm1$   \\
$1024^2$ & $0.0438$    & $155\pm2$      & $6$       & $5$    \\ \bottomrule
\end{tabular}
\caption{$\beta=6$}
\label{tab::Beta6It}
\end{subtable}

\caption{Mean FGMRES iteration count and its deviation when using DD$\alpha$AMG and AM-AMG.
The results were averaged over solutions for 10 distinct configurations, each with 10 right‑hand sides, \textit{i.e.}\ a total of 100 examples for the same $m_0$, $\beta$, and $V$. When the uncertainty is smaller than $10^{-3}$, the deviation is not displayed. The relative tolerance for the solvers is fixed to $10^{-10}$.}
\label{tab::TabIterations}
\end{table*}

\begin{table*}[ht]
\begin{subtable}{\textwidth}
\centering
\begin{tabular}{@{}ccccccc@{}}
\toprule
$V$      & $|m_0-m_c|$ & CG               & Setup   & DD$\alpha$AMG     & Setup  & AM-AMG              \\ \midrule
$128^2$  & $0.0084$    & $24.3\pm2.2$     & $3.7$   & $29.2\pm10.9$     & $0.3$  & $55.0\pm36.9$       \\
$256^2$  & $0.0084$    & $149.4\pm15.5$   & $14.6$  & $134.1\pm36.4$    & $1.1$  & $220.1\pm70.4$      \\
$512^2$  & $0.0084$    & $924.2\pm161.1$  & $58.3$  & $1378.6\pm448.7$  & $4.4$  & $1772.8\pm747.4$    \\ \midrule
$128^2$  & $0.01$      & $22.2\pm2.0$     & $3.7$   & $18.3\pm1.6$      & $0.3$  & $25.4\pm3.5$        \\
$256^2$  & $0.01$      & $172.3\pm17.9$   & $14.6$  & $269.1\pm38.3$    & $1.1$  & $1003.5\pm353.5$    \\
$512^2$  & $0.01$      & $832.1\pm85.2$   & $58.3$  & $1048.5\pm525.3$  & $4.4$  & $2803.3\pm3909.7$   \\
$1024^2$ & $0.01$      & $4442.6\pm328.9$ & $234.1$ & $4903.3\pm1324.9$ & $17.6$ & $11046.7\pm10129.7$ \\ \midrule
$128^2$  & $0.1$       & $3.8\pm0.2$      & $3.6$   & $7.3\pm0.1$       & $0.3$  & $3.1$         \\
$256^2$  & $0.1$       & $16.8\pm1.2$     & $14.3$  & $29.6\pm0.1$      & $1.1$  & $12.5\pm0.5$        \\
$512^2$  & $0.1$       & $67.9\pm4.1$     & $58.3$  & $114.6\pm0.4$     & $4.4$  & $49.4\pm1.3$        \\
$1024^2$ & $0.1$       & $303.5\pm15.9$   & $229.1$ & $475.4\pm1.6$     & $17.6$ & $202.8\pm8.6$       \\ \bottomrule
\end{tabular}
\caption{$\beta=2$}
\label{tab::Beta2Flops}
\end{subtable}

\vspace{0.5 mm}

\begin{subtable}{\textwidth}
\centering
\begin{tabular}{@{}ccccccc@{}}
\toprule
$V$      & $|m_0-m_c|$ & CG               & Setup   & DD$\alpha$AMG    & Setup  & AM-AMG           \\ \midrule
$128^2$  & $0.001$     & $26.1\pm1.5$     & $3.7$   & $25.4\pm2.8$     & $0.3$  & $40.2\pm5.9$     \\
$256^2$  & $0.001$     & $190.2\pm23.5$   & $14.6$  & $105.3\pm8.7$    & $1.1$  & $164.5\pm34.2$   \\
$512^2$  & $0.001$     & $767.4\pm33.7$   & $58.3$  & $606.3\pm38.9$   & $4.4$  & $645.7\pm85.0$   \\
$1024^2$ & $0.001$     & $3431.4\pm295.6$ & $234.0$ & $1510.1\pm83.5$  & $17.6$ & $1900.5\pm211.1$ \\ \midrule
$128^2$  & $0.01$      & $18.7\pm1.6$     & $3.7$   & $15.2\pm0.7$     & $0.3$  & $18.5\pm1.5$     \\
$256^2$  & $0.01$      & $125.1\pm14.1$   & $14.6$  & $79.8\pm5.0$     & $1.1$  & $105.7\pm11.7$   \\
$512^2$  & $0.01$      & $621.4\pm67.6$   & $58.3$  & $406.5\pm33.1$   & $4.4$  & $396.3\pm43.8$   \\
$1024^2$ & $0.01$      & $2840.3\pm496.0$ & $233.9$ & $1347.6\pm153.1$ & $17.6$ & $1575.6\pm263.6$ \\ \midrule
$128^2$  & $0.1$       & $3.0\pm0.1$      & $3.5$   & $7.0\pm0.2$      & $0.3$  & $2.9\pm0.1$      \\
$256^2$  & $0.1$       & $12.8\pm0.3$     & $14.0$  & $29.0\pm0.6$     & $1.1$  & $11.1$     \\
$512^2$  & $0.1$       & $48.8\pm0.8$     & $58.3$  & $106.6\pm2.3$    & $4.4$  & $43.0\pm2.0$     \\
$1024^2$ & $0.1$       & $149.3\pm23.7$   & $216.0$ & $403.9\pm25.7$   & $17.6$ & $142.2\pm15.1$   \\ \bottomrule
\end{tabular}
\caption{$\beta=4$}
\label{tab::Beta4Flops}
\end{subtable}

\vspace{0.5 mm}

\begin{subtable}{\textwidth}
\centering
\begin{tabular}{@{}ccccccc@{}}
\toprule
$V$      & $|m_0-m_c|$ & CG               & Setup   & DD$\alpha$AMG    & Setup  & AM-AMG           \\ \midrule
$128^2$  & $0.001$     & $29.7\pm1.2$     & $3.7$   & $24.1\pm1.7$     & $0.3$  & $38.3\pm4.9$     \\
$256^2$  & $0.001$     & $181.4\pm18.3$   & $14.6$  & $103.4\pm7.4$    & $1.1$  & $145.7\pm14.2$   \\
$512^2$  & $0.001$     & $835.1\pm30.4$   & $58.3$  & $555.2\pm21.4$   & $4.4$  & $584.7\pm46.8$   \\
$1024^2$ & $0.001$     & $4604.5\pm654.4$ & $233.8$ & $1947.9\pm190.4$ & $17.6$ & $2550.6\pm416.4$ \\ \midrule
$128^2$  & $0.01$      & $21.0\pm1.2$     & $3.6$   & $15.6\pm0.4$     & $0.3$  & $18.2\pm1.2$     \\
$256^2$  & $0.01$      & $135.6\pm19.1$   & $14.6$  & $76.4\pm5.9$     & $1.1$  & $95.2\pm11.1$    \\
$512^2$  & $0.01$      & $522.8\pm58.4$   & $58.3$  & $346.5\pm18.3$   & $4.4$  & $307.0\pm21.9$   \\
$1024^2$ & $0.01$      & $2539.8\pm274.3$ & $233.5$ & $1183.9\pm80.6$  & $17.6$ & $1297.2\pm141.8$ \\ \midrule
$128^2$  & $0.0438$    & $6.3\pm0.2$      & $3.6$   & $9.0$      & $0.3$  & $5.4\pm0.1$      \\
$256^2$  & $0.0438$    & $27.4\pm1.1$     & $14.4$  & $36.2\pm0.1$     & $1.1$  & $20.9\pm0.7$     \\
$512^2$  & $0.0438$    & $73.2\pm13.6$    & $58.3$  & $120.0\pm8.1$    & $4.4$  & $55.2\pm8.2$     \\
$1024^2$ & $0.0438$    & $66.2\pm1.0$     & $169.7$ & $277.7\pm1.1$    & $17.6$ & $106.0\pm0.1$    \\ \bottomrule
\end{tabular}
\caption{$\beta=6$}
\label{tab::Beta6Flops}
\end{subtable}

\caption{Mean number of GFLOP count and its deviation. The GFLOP count is also determined for the setup phase. The operation numbers reported for the solvers include the contribution from setup phase. The results were averaged over solutions for 10 distinct configurations, each with 10 right‑hand sides, \textit{i.e}.\ a total of 100 examples for the same $m_0$, $\beta$, and $V$. When the uncertainty is smaller than $10^{-3}$, the deviation is not displayed. The relative tolerance for the solvers is fixed to $10^{-10}$.}
\label{tab::TabFlops}
\end{table*}

\FloatBarrier
\section{Conclusion}\label{conclusions}

A comparison of the efficiency of two multilevel preconditioners for the Dirac matrix in the Schwinger model was done. Close to criticality and on fine lattices, $\beta=4,6$, both methods outperform Conjugate Gradient and alleviate critical slowing down. However, they have fundamental differences. DD$\alpha$AMG relies on test vectors for the interpolation scheme. To have better convergence, the amount of these vectors must be adapted depending on the problem size. No such requirement exists for AM-AMG. Once the problem size, preconditioner, and accelerator (outer solver) have been fixed, AM-AMG can be applied across different volumes and $\beta$ values without further adjustment. \\

Our results show that AM-AMG might be a competent method for computing quark propagators close to physical masses. Although it does not generally outperform DD$\alpha$AMG, it offers greater flexibility across varying scenarios and its setup phase is substantially less expensive, as shown in Table \ref{tab::TabFlops}. Furthermore, the current implementation is not specifically optimized for Lattice Field Theory, leaving a lot of room for improvement. \\

In the future, we want to develop SM-AMG for QCD in a complex form that does not rely on the K-formulation nor SAMG and therefore needs less memory. Additionally, we plan to incorporate the underlying geometry in the SM-AMG approach. In Refs.\ \cite{ehrmannAMAMG,dissAMAMG} one can see that the SM-AMG approach tends to have better convergence than AM-AMG. With this, we hope to achieve a faster solver and better convergence. Another possible improvement could involve incorporating test vectors into the SM-AMG method to better capture the near-kernel (augmented Krylov methods). However, this would inevitably increase the cost of the setup phase.

\section{Data and software availability}
Data are available upon request. The HMC and DD$\alpha$AMG implementations used in this work are available on \cite{hmc,ddalpha}. The tests of AM-AMG are performed with Fraunhofer's SAMG library version 2024 \cite{SAMG}.   

\section{Acknowledgments}\label{acknowledgments}

We would like to thank Travis Whyte for reading the manuscript and providing valuable comments. We also thank Edgar Landinez and Bartosz Kostrzewa for their assistance with profiling tools. We acknowledge funding from the Deutsche Forschungsgemeinschaft (DFG, German Research Foundation) as part of the CRC 1639 NuMeriQs project number 511713970. The HMC for generating configurations and DD$\alpha$AMG runs were performed on the supercomputer JURECA (JSC), with computing time granted from the SDL Numerical Quantum Field Theory group.\\

\clearpage
\printbibliography[title={References}]

\typeout{get arXiv to do 4 passes: Label(s) may have changed. Rerun}

\end{document}